\documentclass[final]{elsart}
\usepackage[T1]{fontenc}
\usepackage[latin9]{inputenc}
\usepackage{amsmath}
\usepackage{graphicx}
\usepackage{amsfonts}
\usepackage[english]{babel}

\journal{PRD}
\begin{document}

\begin{frontmatter}

\title{Kaluza-Klein theory in the limit of large number of extra dimensions}

\author{Fabrizio Canfora}

\address{Centro de Estudios Cient\'ificos (CECS), Valdivia, Chile}

\ead{canfora@cecs.cl}

\author{Alex Giacomini}
\address{Instituto de F\'isica, Facultad de Ciencias, Universidad Austral de
Chile, Casilla 567, Valdivia,Chile}
\ead{alexgiacomini@uach.cl}

\author{Alfonso R. Zerwekh}
\address{Centro de Estudios Subat\'omicos and Instituto de F\'isica, Facultad de Ciencias, Universidad Austral de
Chile, Casilla 567, Valdivia,Chile}
\ead{alfonsozerwekh@uach.cl}





\date{}
\begin{abstract}
The Kaluza-Klein compactification in the limit of large number of extra dimensions is studied. Starting point is the 
Einstein-Hilbert action plus cosmological constant in 4+D dimensions. It is shown that in the large D limit the effective 
four dimensional cosmological constant is of order $1/D$ whereas the size of the extra dimensions remains finite. 
A 't Hooft like large D expansion of the effective Lagrangian for the Kaluza-Klein scalar and gauge fields arising from 
the dimensional reduction is considered. It is shown that the propagator of the scalar field associated to the 
determinant of the metric of the extra dimensions is strongly suppressed. This is an interesting result as in standard 
Kaluza-Klein theory  this scalar degree of freedom is responsible for the constraint on the gauge fields which makes 
it impossible to recover the usual  Yang-Mills equations. Moreover in the large D limit it turns out that the ultraviolet 
divergences due to the interactions between gauge and scalar fields
are softened. 
\end{abstract}
\end{frontmatter}

\section{Introduction}

\noindent The Kaluza-Klein scenario aiming to recover gauge fields from pure
space-time geometry is one of the most fascinating ideas of theoretical
physics. In the original proposal, the attempt was to unify four-dimensional
General Relativity (GR) with Maxwell electrodynamics as GR in five
dimensions compactified on a circle. In this way one gets Einstein's
equations in four dimensions, a gauge field and also a scalar field.
However, a quite serious problem in trying to make contact with gauge theory
arises. The dynamics of Maxwell field is not exactly what one would like
because the extra scalar degree of freedom (which corresponds to the
determinant of the metric along the extra-dimension) gives rise to an extra
constraint which prevents one from having both a constant scalar field and
the usual Maxwell equations for the gauge field.

In order to include non-Abelian gauge fields the curvature of the extra
dimensions cannot vanish. This is problematic since the product of four
dimensional Minkowski with a compact manifold with non-Abelian isometry
group $G$ (the natural ground state of Kaluza-Klein compactification) is not
a solution of higher dimensional GR \cite{Appelquist:1983vs}. A
non-vanishing positive cosmological constant may help since solutions which
are the product of a four dimensional Lorentzian manifold of constant
positive curvature with a compact manifold with non-Abelian isometry group $%
G $ exist. However in this case the effective four-dimensional cosmological
constant turns out to be of the same order of magnitude as the curvature of
the compact space. It is therefore difficult to make contact with
phenomenology if one assumes that the compact extra-dimensions are
characterized by a scale much smaller than the macroscopic four dimensions
(a bright analysis of the problem of Kaluza-Klein compactification is in 
\cite{Witten:1981me}; for updated reviews, see, e.g., \cite{Blau:1986mi}, 
\cite{KKR}, \cite{DK07}). Also the problem already mentioned above remains:
namely, the dynamics of Yang-Mills field is not exactly what one would like
because the extra scalar degrees of freedom (and in particular, the degree
of freedom corresponding to the determinant of the metric along the
extra-dimension) give rise to an extra constraints absent in Yang-Mills
theory. It has been recently shown \cite{CGTW} that in the context of
Lovelock gravities many of the problem of usual Kaluza-Klein
compactifications can be addressed.

One may also be interested in analyzing the quantum features of the
effective Kaluza-Klein Lagrangian for scalar and gauge fields (thinking at
the four-dimensional metric as a classical background on which the
Kaluza-Klein scalars and the gauge fields propagates\footnote{%
This makes sense if the typical length scale of the extra dimensions is much
smaller than the typical length scale of the macroscopic four-dimensional
metric.}). Indeed, such Lagrangian contains non-renormalizable interactions
between the scalars and the gauge fields which generate many problems in the
Ultra-Violet (UV) limit.

Here, we propose a framework in which the problems above described can be
treated in a natural way: we will apply the well known 't Hooft large 
\textbf{N}\ expansions \cite{T74a} \cite{T74b}\footnote{%
The Veneziano limit \cite{Ve76}, in which the ratio \textbf{N/N}$_{f}$ is
kept fixed (\textbf{N}$_{f}$ being the number of quarks\ flavours), was also
important to further clarify several features of quark and mesons dynamics;
two pedagogical reviews are \cite{Ma98}.} to the effective Kaluza-Klein
Lagrangian for scalars and gauge fields (thus, in the present case, \textbf{N%
}\ will be related to the number $\mathbf{D}$\ of extra-dimensions).

The first attempts to obtain an expansion similar to the 't Hooft one in
gravity have been performed in \cite{To77}\ \cite{Str81} \cite{Smo82} and
further refined in \cite{Bo03}. In \cite{Str81} and in \cite{Bo03} the
"{}{}small parameter"{} is $1/\mathbf{d}$ ($\mathbf{d}$\ being the total
number of space-time dimensions). While in \cite{Ca1} \cite{Ca2}, it has
been proposed to think at the (Euclidean) four-dimensional GR as a
constrained gauge theory for the $\mathbf{SO(4)}$ group and performs an
expansion in which $\mathbf{4\rightarrow N}$\ is large (keeping fixed the
number of spacetime dimensions, that is $\mathbf{d}=4$).

For the Kaluza-Klein compactification of GR plus cosmological constant in 4+$%
\mathbf{D}$ dimensions, the subject of the present paper, it is found that
already at classical level there is a nontrivial large $\mathbf{D}$
expansion whose most remarkable feature is that the effective four
dimensional cosmological constant is of order of $1/\mathbf{D}$. At quantum
level, performing the 't Hooft like large $\mathbf{D}$ expansion, it is
found that the propagator of the scalar field corresponding to the
determinant of the metric of the extra dimensions is strongly suppressed.
This is a nontrivial feature as in standard Kaluza-Klein theory such scalar
degree of freedom is responsible for the extra constraint on the gauge
fields mentioned above which makes it impossible to recover the Yang-Mills
equations when this field is constant.

Moreover, from the "{}large \textbf{N}"{} perspective, the Kaluza-Klein
effective Lagrangian presents new features which are absent in the large 
\textbf{N} expansion of QCD or in the large \textbf{N} expansions of model
with global symmetries (such as the Gross-Neveu model; for two reviews see 
\cite{Ma98}). These novel features allow one to soften the UV problems
already mentioned\footnote{%
This framework is somehow in between the points of view of the references 
\cite{Str81} and \cite{Bo03} (in which a large $d$ expansion was considered)
and the point of view proposed in \cite{Ca1} \cite{Ca2} (where $d$ is kept
fixed and "{}{}$\mathbf{SO(4)}$ is enlarged"{}{} in such a way to separate
the "{}{}internal indices"{}{} from the "{}{}space-time indices"{}{}).}.

Indeed, already the classical theory manifests a non-trivial large $\mathbf{D}$ scaling, so one could wonder about the justification of treating $1/\mathbf{D}$ as a coupling constant. On the other hand, as it is well known, in quantum mechanics one reaches the semi-classical regime in the limit of very high quantum numbers. In quantum field theory, the semi-classical regime is valid when the vacuum expectation value(s) of the number operator(s) of the field(s) is (are) very large (as it happens, for instance, when condensates appear). Therefore, since in the large $\mathbf{D}$ expansion the number of degrees of freedom grows polynomially with $\mathbf{D}$, one can treat $1/\mathbf{D}$ as a small parameter for the semi-classical expansion around the Kaluza-Klein vacuum (in analogy with what happens in the large $\mathbf{N}$ expansion of the 3D Gross-Neveu model). In fact, as in quantum field theory condensates break some symmetry of the theory, the Kaluza-Klein vacuum breaks part of the symmetry of the trivial maximally symmetric vacuum. Furthermore, as it will be shown in the next sections, the large $\mathbf{D}$ expansion (at least partially) solves some consistency problems of the classical Kaluza-Klein theory.

The structure of the paper is the following: First the classical nontrivial
features of the Large $\mathbf{D}$ limit of Kaluza-Klein compactification
which arise at classical level are discussed. Then the basic features of the
't Hooft expansion and of some nontrivial large $\mathbf{D}$-resummations
are discussed. It is found that the two most remarkable features of this
expansion are the suppression of the scalar degree of freedom corresponding
to the determinant of the metric and the softening of the ultraviolet
divergences. In the last sections the conclusions are presented.

\section{The Kaluza-Klein scenario: a short introduction}

Let us consider the Kaluza-Klein scenario in $\left( 4+\mathbf{D}\right) $
dimensions whose ground state is a product manifold $M_{4}\times K_{\mathbf{D%
}}$ ($K_{\mathbf{D}}$ being an Euclidean manifold of constant positive
curvature; nice reviews on this subject are \cite{KKR}, \cite{Mar84}, \cite%
{Blau:1986mi}). Here, we will only consider the Einstein-Hilbert action with
a positive cosmological constant $\Lambda $,%
\begin{equation*}
\Lambda _{4+\mathbf{D}}=\Lambda ,
\end{equation*}%
(in order to have non-Abelian gauge fields) in $\left( 4+\mathbf{D}\right) $
dimensions. The ground state metric is the following direct product in which
the extra-dimensional manifold is a constant curvature manifold%
\begin{equation*}
g_{(4+\mathbf{D})}=g_{\mu \nu }\left( x^{\mu }\right) dx^{\mu }dx^{\nu }+%
\widehat{g}_{ab}\left( y^{a}\right) dy^{a}dy^{b}
\end{equation*}%
where the coordinates $x^{\mu }$ are intrinsic to $M_{4}$ and $y^{a}$ are $%
\mathbf{D}$-dimensional coordinates intrinsic to $K_{\mathbf{D}}$. With this
ansatz, the mixed components of the Einstein equation (written in the usual
second order formalism\footnote{%
In the next subsection, the Palatini first order formalism will be
considered.}) involving the mixed components $G_{a\mu }$ of the $\left( 4+%
\mathbf{D}\right) $-dimensional Einstein tensor $G_{AB}$ are trivially
satisfied. The $\left( 4+\mathbf{D}\right) $ dimensional Einstein equations%
\begin{equation*}
G_{AB}=\Lambda g_{AB}
\end{equation*}%
reduce to a four-dimensional Einstein equations for $g_{\mu \nu }$: 
\begin{equation*}
G_{\mu \nu }^{(4)}+\Lambda _{4}g_{\mu \nu }=0\,,
\end{equation*}%
with an effective cosmological constant $\Lambda _{4}$ and to a $\mathbf{D}$%
-dimensional Euclidean Einstein equations for $\widehat{g}_{ab}$:%
\begin{equation*}
G_{ab}^{(\mathbf{D})}+\Lambda _{\mathbf{D}}g_{ab}=0\,.
\end{equation*}%
As it is well known, a non-Abelian algebra of Killing fields is only
compatible with a $\mathbf{D}-$dimensional symmetric space of positive
effective cosmological constant $\Lambda _{\mathbf{D}}$: the effective four
dimensional and $\mathbf{D}-$dimensional cosmological constants are
respectively 
\begin{equation*}
\Lambda _{4}=\frac{2}{\mathbf{D}+2}\Lambda 
\end{equation*}%
and 
\begin{equation*}
\Lambda _{\mathbf{D}}=\frac{\mathbf{D}-2}{\mathbf{D}+2}\Lambda 
\end{equation*}%
so that%
\begin{equation*}
\frac{\Lambda _{4}}{\Lambda _{\mathbf{D}}}=\frac{2}{\mathbf{D}-2}.
\end{equation*}%
Therefore, when $\mathbf{D}$ is large, $\Lambda _{4}${} is much smaller than 
$\Lambda _{\mathbf{D}}$: in particular, at leading order in the large $%
\mathbf{D}$\ expansion, $\Lambda _{4}$ vanishes. One then see that the large 
$\mathbf{D}$ expansion itself is able to keep separated the macroscopic
four-dimensional scale from the typical extra-dimensional scale without any
extra ingredient. This is indeed a very attractive feature of the present
framework. In the following subsection, the non-trivial large $\mathbf{D}$
scaling of the classical theory will be deduced in the first order Palatini
formalism which, for this goal, is more convenient than the second order
formalism.

\subsection{Kaluza-Klein scenarios in the first order formalism}

To fully display the $\mathbf{D}$-dependence in the large $\mathbf{D}$
expansion it is convenient to introduce the following notations (we will
follow \cite{Mar84}): let $t_{a}$ be the Lie algebra generators
corresponding to the Lie group $G$ (which will play the role of the gauge
group of the Kaluza-Klein gauge fields):

\begin{eqnarray*}
\left[ t_{a},t_{b}\right] &=&C_{ab}^{c}t_{c},\ \ \ s^{-1}ds=e^{a}t_{a},\  \\
dss^{-1} &=&-\widehat{e}^{a}t_{a}, \\
s &\in &G,\ \ \ \ e^{a}\left( Y_{b}\right) =\delta _{b}^{a},\ \ \ \ \ \ \ \ 
\widehat{e}^{a}\left( \widehat{Y}_{b}\right) =\delta _{b}^{a}
\end{eqnarray*}%
here $\widehat{Y}_{a}$ and $Y_{b}$\ represent the right and left invariant
vector fields while $\widehat{e}^{a}$ and $e^{a}$ are the corresponding dual
one-forms. To have a consistent Kaluza-Klein scenario one may consider the
case in which the $Y_{a}$ are the Killing vectors of the full $(4+\mathbf{D}%
) $-dimensional metric\footnote{%
It can be also analyzed the case in which the $Y_{a}$ are the Killing
vectors of the metric only when restricted to the extra-dimensions \cite%
{Mar84}, but we will restrict the present analysis only to the case in which
the $Y_{a}$ are the Killing vectors of the total metric.}. The natural
Kaluza-Klein ground state is a product of a four dimensional manifold
fulfilling the four-dimensional Einstein equations (with a small
cosmological constant if $\mathbf{D}$ is large) times a coset manifold $G/H$
invariant under the corresponding non-Abelian algebra of Killing fields. The
ground state metric (whose Killing vectors are the $Y_{a}$) on $G/H$\ will
be written as 
\begin{equation}
g_{G/H}=\widehat{g}_{ab}\widehat{e}^{a}\widehat{e}^{b}.  \label{gs1}
\end{equation}%
In the ground state, $\widehat{g}_{ab}$ does not depend on $x$: to fix the
idea, one can think at the extra-dimensional manifold corresponding to the
ground state of the Kaluza-Klein scalars as the $\mathbf{D}$-sphere%
\begin{equation}
\frac{G}{H}=S^{\mathbf{D}}=\frac{SO(\mathbf{D}+1)}{SO(\mathbf{D})}.
\label{gs2}
\end{equation}%
At the semi-classical level, the large $\mathbf{D}$ expansion corresponds,
from the point of view of gauge fields, to a (bit unusual as we shall see in
the next section) 't Hooft expansion of the effective Kaluza-Klein
Largrangian for scalars and gauge fields with $SO(\mathbf{D})$\ as gauge
group.

Eventually, the usual Kaluza-Klein ansatz for the $\left( 4+\mathbf{D}%
\right) $-dimensional metric $g_{(4+\mathbf{D})}$ reads%
\begin{equation}
g_{(4+D)}=g_{\mu \nu }dx^{\mu }dx^{\nu }+\widehat{g}_{ab}\left( x^{\mu
}\right) \left( \widehat{e}^{a}+A^{a}\right) \left( \widehat{e}%
^{b}+A^{b}\right) .  \label{kakame}
\end{equation}%
The above metric (\ref{kakame})\ is left unchanged by the following gauge
transformations:%
\begin{equation*}
A^{\prime }=u^{-1}Au+du^{-1},\ \ \ \ \left( \widehat{g}_{ab}\right) ^{\prime
}=\left( \widehat{g}_{cd}\right) R\left( u\right) _{a}^{c}R\left( u\right)
_{b}^{d}
\end{equation*}%
where $u\left( x^{\mu }\right) \in SO(D)$, while the matrix $R\left(
u\right) _{a}^{c}$ is in the adjoint representation in the sense that the
element $u\in SO(D)$ induces the following transformation on generators $%
t_{a}$%
\begin{equation*}
\left( t_{a}\right) ^{\prime }=t_{b}R\left( u\right) _{a}^{b}.
\end{equation*}%
As a consequence, since in the 't Hooft notation the propagators of the "{}{}%
$SO(\mathbf{D})$ gluons"{}{} $A^{b}$ are represented by a double line (as
usual, the $A^{a}$ fields transform in the adjoint) the scalars degrees of
freedom corresponding to $\widehat{g}_{ab}\left( x^{\mu }\right) $ (see Eq. (%
\ref{pidef}) below) will be represented by four lines as it will be
explained in more details in the next section (a similar phenomenon also
occurs in \cite{Ca1} \cite{Ca2}): this is the origin of the unusual features
of the 't Hooft expansion of the Kaluza-Klein effective Lagrangian.

In many field theoretical models in which the large \textbf{N} expansion is
available (such as the Gross-Neveu, Yang-Mills theory, and so on) the
non-trivial scaling with \textbf{N} only appears at a quantum level (see,
for instance, \cite{Ma98}). Namely, only after computing Feynman diagrams
with loops, one can recognize the possibility to perform a large \textbf{N}
expansion which corresponds to a semi-classical expansion. However, in
gravity (because of the fact that the extra-dimensions describe in a sense
the local gauge symmetry of Yang-Mills theory), already the classical
equations of motion manifest a non-trivial scaling with $\mathbf{D}$
(related to the dimension of the Kaluza-Klein gauge fields).

Let $\omega ^{AB}$ and $e^{A}$ the (torsion free) spin connection and the
"{}{}$(4+\mathbf{D})-$bein"{}{} respectively, the Riemann curvature two form 
$R^{AB}$ is defined as 
\begin{eqnarray*}
R^{AB} &=&d\omega ^{AB}+\omega _{\ C}^{A}\omega ^{CB},\ \ \ \ A,B,C,..=1,..,%
\mathbf{D}+4,\ \ \ \mu ,\nu ,\rho ,\sigma =1,..,4, \\
De^{A} &=&T^{A}=de^{A}+\omega _{\ C}^{A}e^{C}=0,\ \ \ \
a,b,c,...,a_{1},a_{2},...,i,j,k,...=1,..,\mathbf{D}.
\end{eqnarray*}%
The Einstein-Hilbert action plus the cosmological term in $4+\mathbf{D}$
dimensions in the first order formalism read%
\begin{equation}
I_{\mathbf{D}+4}={\displaystyle\int \left( \frac{c_{0}}{4+\mathbf{D}}%
e^{A_{1}}..e^{A_{4+D}}+\frac{c_{1}}{\mathbf{D}+2}%
R^{A_{1}A_{2}}e^{A_{3}}..e^{A_{4+D}}\right) \label{act1}}
\end{equation}%
where $c_{0}$ is proportional to the cosmological constant and $c_{1}$ to
the $4+\mathbf{D}$ Newton constant. At a classical level, no argument can be
invoked which suggests that $c_{0}$ is negligible with respect to $c_{1}$ so
that the "bare" classical coupling constants $c_{0}$ and $c_{1}$ scale with $%
\mathbf{D}$ in the same way: therefore, in the large $\mathbf{D}$ limit, $%
c_{0}/c_{1}$ is a non-vanishing finite constant (let us call such a constant 
$\frac{6\widehat{\beta }}{\widehat{\alpha }}$ for future convenience)%
\begin{equation*}
\frac{c_{0}}{c_{1}}\underset{\mathbf{D}\gg 1}{\approx }\frac{6\widehat{\beta 
}}{\widehat{\alpha }}+o(1/\mathbf{D}).
\end{equation*}

Let us divide the indices into two groups: $\mu $, $\nu $, $\rho $, $\sigma $%
,... represent the macroscopic Lorentzian four dimensions ($\mu $, $\nu $, $%
\rho $, $\sigma =1,..,4$) while small Latin indices $a$, $b$, $c$,..$i$,$j$,$%
k$,.., $a_{1}$, $a_{2}$,..\ (which will play the role of the internal
indices of the Yang-Mills fields) represent the $\mathbf{D}$ compact extra
dimensions ($a$, $b$, $c$,..$i$,$j$,$k$,.., $a_{1}$, $a_{2}$,.. $=1,..,%
\mathbf{D}$). Thus, there are three different kinds of components of the
Riemann curvature two form $R^{AB}$:%
\begin{equation*}
R^{\mu \nu },\ \ \ \ R^{\mu a},\ \ \ \ R^{ab}.
\end{equation*}%
Roughly speaking, the components $R^{\mu \nu }$\ give rise to the usual four
dimensional gravitational interaction of GR with a suitable energy-momentum
tensor for the gauge and scalar fields as source, the components $R^{\mu a}$%
\ are related to the field strength of the gauge fields (generating the
corresponding equations of motion) while the components $R^{ab}$ are related
to the scalar fields and to the well known scalar constraint on the gauge
fields (the explicit decomposition of the Riemann tensor can be found, for
instance, in \cite{KKR}, \cite{Mar84}, \cite{Blau:1986mi}). The equations of
motion corresponding to the action in Eq. (\ref{act1}) split as follows:%
\begin{eqnarray*}
0 &=&E_{\mu }=\varepsilon _{\mu \nu \rho \sigma a_{1}..a_{\mathbf{D}%
}}\left\{ c_{0}\frac{\left( \mathbf{D}+3\right) \left( \mathbf{D}+2\right) }{%
6}\left( e^{\nu }e^{\rho }e^{\sigma }e^{a_{1}}..e^{a_{\mathbf{D}}}\right)
+\right. \\
&&+c_{1}\mathbf{D}\left( \mathbf{D}-1\right) \left[ \frac{R^{a_{1}a_{2}}}{6}%
\left( e^{\nu }e^{\rho }e^{\sigma }e^{a_{3}}..e^{a_{\mathbf{D}}}\right)
+\right.
\end{eqnarray*}%
\begin{equation*}
\left. \left. +\frac{R^{\nu a_{1}}}{\mathbf{D}-1}\left( e^{\rho }e^{\sigma
}e^{a_{2}}..e^{a_{\mathbf{D}}}\right) +\frac{\left( e^{a_{1}}..e^{a_{\mathbf{%
D}}}e^{\sigma }\right) }{\left( \mathbf{D}-1\right) \mathbf{D}}R^{\nu \rho }%
\right] \right\} ,
\end{equation*}%
\begin{eqnarray*}
0 &=&E_{a_{1}}=\varepsilon _{\mu \nu \rho \sigma a_{1}..a_{\mathbf{D}%
}}\left\{ c_{0}\frac{\left( \mathbf{D}+3\right) \left( \mathbf{D}+2\right) }{%
24}\left( e^{\mu }e^{\nu }e^{\rho }e^{\sigma }e^{a_{2}}..e^{a_{\mathbf{D}%
}}\right) +\right. \\
&&+c_{1}\left( \mathbf{D}-2\right) \left( \mathbf{D}-1\right) \left[ \frac{%
R^{a_{2}a_{3}}}{24}\left( e^{\mu }e^{\nu }e^{\rho }e^{\sigma
}e^{a_{4}}..e^{a_{\mathbf{D}}}\right) +\right.
\end{eqnarray*}%
\begin{equation*}
\left. \left. +\frac{R^{\mu a_{2}}}{3\left( \mathbf{D}-2\right) }\left(
e^{\nu }e^{\rho }e^{\sigma }e^{a_{3}}..e^{a_{\mathbf{D}}}\right) +\frac{%
\left( e^{\rho }e^{\sigma }e^{a_{2}}..e^{a_{\mathbf{D}}}\right) }{2\left( 
\mathbf{D}-1\right) \left( \mathbf{D}-2\right) }R^{\mu \nu }\right] \right\}
.
\end{equation*}%
It then clear that when $\mathbf{D}$ is very large the above equations
separate into decoupled equations for the different components $R^{\mu \nu }$%
, $R^{\mu a}$ and $R^{ab}$: the reason is that for large $\mathbf{D}$ the
number of scalar field components grows faster than the number of
Kaluza-Klein gauge fields while the number of four dimensional gravitational
degrees of freedom does not change.

We will assume, as it is usually done in various types of large \textbf{N}
expansions, that for very large $\mathbf{D}$ any field $\Phi $ can be
expanded as follows%
\begin{equation*}
\Phi =\Phi _{\left( 0\right) }+\frac{1}{\mathbf{D}}\Phi _{\left( 1\right) }+%
\frac{1}{\mathbf{D}^{2}}\Phi _{\left( 2\right) }+...,
\end{equation*}%
\begin{equation*}
such\ \ \ that,\ \ \ \ \forall \ \ \ \ k,\ \ \ \Phi _{\left( k\right) }\ \ \
does\ \ \ not\ \ \ depend\ \ \ on\ \ \ \mathbf{D},
\end{equation*}%
where $\Phi _{\left( 0\right) }$ is the leading order and the terms $\Phi
_{\left( i\right) }$ for $i>0$ can be considered as subleading corrections
so that no component of $R^{AB}$ is divergent at large $\mathbf{D}$. Indeed,
such an hypothesis is the most natural one since the large $\mathbf{D}$
expansion itself provides one with a suitable tool to keep well separated
the macroscopic scale of the four dimensional directions ($\mu $, $\nu $%
,...) from the compactified directions ($a_{1}$, $a_{2}$,...).

To simplify the notation, it is convenient to define two rescaled coupling
constants $\widehat{\beta }$\ and $\widehat{\alpha }$\ in terms of $c_{0}$
and $c_{1}$ as follows:%
\begin{equation*}
c_{0}=\frac{6\widehat{\beta }}{\left( \mathbf{D}+3\right) \left( \mathbf{D}%
+2\right) },\ \ \ c_{1}=\frac{\widehat{\alpha }}{\mathbf{D}\left( \mathbf{D}%
-1\right) }.
\end{equation*}%
The field equations $E_{\mu }=0$ and $E_{a_{1}}=0$ now read%
\begin{eqnarray}
0 &=&E_{\mu }=\varepsilon _{\mu \nu \rho \sigma a_{1}..a_{\mathbf{D}%
}}\left\{ \widehat{\beta }\left( e^{\nu }e^{\rho }e^{\sigma
}e^{a_{1}}..e^{a_{\mathbf{D}}}\right) +\right.   \label{eqkk1} \\
&&+\left. \widehat{\alpha }\left[ \frac{R^{a_{1}a_{2}}}{6}\left( e^{\nu
}e^{\rho }e^{\sigma }e^{a_{3}}..e^{a_{\mathbf{D}}}\right) +\frac{R^{\nu
a_{1}}}{\mathbf{D}-1}\left( e^{\rho }e^{\sigma }e^{a_{2}}..e^{a_{\mathbf{D}%
}}\right) +\frac{\left( e^{a_{1}}..e^{a_{\mathbf{D}}}e^{\sigma }\right) }{%
\left( \mathbf{D}-1\right) \mathbf{D}}R^{\nu \rho }\right] \right\} ,  \notag
\end{eqnarray}
\begin{eqnarray}
0 &=&E_{a_{1}}=\varepsilon _{\mu \nu \rho \sigma a_{1}..a_{\mathbf{D}%
}}\left\{ \frac{\widehat{\beta }}{4}\left( e^{\mu }e^{\nu }e^{\rho
}e^{\sigma }e^{a_{2}}..e^{a_{\mathbf{D}}}\right) +\widehat{\alpha }\left( 1-%
\frac{2}{\mathbf{D}}\right) \cdot \right.   \label{eqkk2} \\
&&\cdot \left[ \frac{R^{a_{2}a_{3}}}{24}\left( e^{\mu }e^{\nu }e^{\rho
}e^{\sigma }e^{a_{4}}..e^{a_{\mathbf{D}}}\right) +\frac{R^{\mu a_{2}}}{%
3\left( \mathbf{D}-2\right) }\left( e^{\nu }e^{\rho }e^{\sigma
}e^{a_{3}}..e^{a_{\mathbf{D}}}\right) +\right.   \notag
\end{eqnarray}%
\begin{equation*}
\left. \left. +\frac{\left( e^{\rho }e^{\sigma }e^{a_{2}}..e^{a_{\mathbf{D}%
}}\right) }{2\left( \mathbf{D}-1\right) \left( \mathbf{D}-2\right) }R^{\mu
\nu }\right] \right\} .
\end{equation*}%
It is easy to see a very nice feature of the present large $\mathbf{D}$
framework: a\textit{\ priori}, one should assume that all the components of
the full Riemann tensor $R^{\mu \nu }$,\ $R^{\mu a}$,\ $R^{ab}$ have already
at a classical level a non-trivial $1/\mathbf{D}$ expansion:%
\begin{eqnarray*}
R^{\mu \nu } &=&R_{\left( 0\right) }^{\mu \nu }+\frac{1}{\mathbf{D}}%
R_{\left( 1\right) }^{\mu \nu }+..., \\
R^{\mu a} &=&R_{\left( 0\right) }^{\mu a}+\frac{1}{\mathbf{D}}R_{\left(
1\right) }^{\mu a}+..., \\
R^{ab} &=&R_{\left( 0\right) }^{ab}+\frac{1}{\mathbf{D}}R_{\left( 1\right)
}^{ab}+....
\end{eqnarray*}%
However, as far as $R^{\mu \nu }$ and\ $R^{\mu a}$\ are concerned, it is
consistent with the field equations to simply consider the leading\ terms:%
\begin{eqnarray*}
R^{\mu \nu } &=&R_{\left( 0\right) }^{\mu \nu }, \\
R^{\mu a} &=&R_{\left( 0\right) }^{\mu a},
\end{eqnarray*}%
while as far as $R^{ab}$\ is concerned it is enough to consider the first
two terms of the expansion: 
\begin{equation}
R^{ab}=R_{\left( 0\right) }^{ab}+\frac{1}{\mathbf{D}}R_{\left( 1\right)
}^{ab}.  \label{extracurv1}
\end{equation}%
Thus, at large $\mathbf{D}$, one gets the following decoupled equations for $%
R_{\left( 0\right) }^{a_{1}a_{2}}$, $R_{\left( 1\right) }^{a_{1}a_{2}}$, $%
R^{\nu a_{1}}$ and $R^{\nu \rho }$:%
\begin{equation}
\varepsilon _{a_{1}..a_{\mathbf{D}}}\left( \frac{\widehat{\beta }}{4}%
e^{a_{2}}..e^{a_{\mathbf{D}}}+\widehat{\alpha }\frac{R_{\left( 0\right)
}^{a_{2}a_{3}}}{24}\left( e^{a_{4}}..e^{a_{\mathbf{D}}}\right) \right) =0,
\label{eq1kk}
\end{equation}%
\begin{equation}
\varepsilon _{a_{1}..a_{\mathbf{D}}}\left( e^{a_{4}}..e^{a_{\mathbf{D}%
}}\right) R_{\left( 1\right) }^{a_{2}a_{3}}=0,  \label{eq2kk}
\end{equation}%
\begin{equation}
\varepsilon _{\mu \nu \rho \sigma a_{1}..a_{\mathbf{D}}}\left( e^{\rho
}e^{\sigma }e^{a_{2}}..e^{a_{\mathbf{D}}}\right) R^{\nu a_{1}}=0,
\label{eq3kk}
\end{equation}%
\begin{equation}
\varepsilon _{\mu \nu \rho \sigma }R^{\nu \rho }e^{\sigma }=0,  \label{eq4kk}
\end{equation}

The $R_{\left( 0\right) }^{a_{2}a_{3}}$ components satisfy Euclidean
Einstein equations with an effective $\mathbf{D}$-dimensional cosmological
constant given by $\frac{6\widehat{\beta }}{\widehat{\alpha }}$.
Thus, no matter how large is the actual $(4+\mathbf{D})$-dimensional
cosmological constant, the consistency of the large $\mathbf{D}$ expansion demands that the effective four-dimensional cosmological constant is of order $1/\mathbf{D}$ (indeed, the effective four-dimensional cosmological constant vanishes at leading order in the large D expansion).
 The leading
correction to $R^{ab}$ in the $1/\mathbf{D}$ expansion (namely, $R_{\left(
1\right) }^{ab}$) satisfies Euclidean $\mathbf{D}-$dimensional Einstein
equations with a vanishing cosmological constant. The non-trivial large $%
\mathbf{D}$ scaling present already in the classical equations of motion is
an interesting feature of the present framework.

\section{Propagators and 't Hooft expansion}

In the next section, some large $\mathbf{D}$\ correction to the propagators
of the scalar fields $\rho $ and $\overrightarrow{\pi }$ will be analyzed:
in order to achieve this goal, it is convenient to use the second order
formalism. The $4+\mathbf{D}$ dimensional gravitational action reads%
\begin{equation*}
S_{4+\mathbf{D}}={\displaystyle\int \sqrt{g_{4+\mathbf{D}}}\left( R_{4+%
\mathbf{D}}+2\Lambda _{4+\mathbf{D}}\right) .}
\end{equation*}%
The $4+\mathbf{D}$ dimensional Ricci scalar can be expressed in terms of the
four dimensional Ricci scalar $R_{4}$, the Kaluza-Klein scalars and gauge
fields as follows%
\begin{eqnarray}
R_{4+\mathbf{D}} &=&R_{4}+R_{\mathbf{D}}-\frac{g^{\mu \nu }g^{\alpha \beta }%
}{4}\widehat{g}_{ab}F_{\mu \alpha }^{a}F_{\nu \beta }^{b}-\nabla _{\mu
}\left( tr\left( \widehat{g}^{-1}\nabla ^{\mu }\widehat{g}\right) \right) + 
\notag \\
&&-\frac{1}{4}tr\left( \left( \widehat{g}^{-1}\nabla ^{\mu }\widehat{g}%
\right) \left( \widehat{g}^{-1}\nabla _{\mu }\widehat{g}\right) \right)
\label{larged1} \\
&&-\frac{1}{4}\left( tr\left( \widehat{g}^{-1}\nabla ^{\mu }\widehat{g}%
\right) \right) \left( tr\left( \widehat{g}^{-1}\nabla _{\mu }\widehat{g}%
\right) \right)  \notag
\end{eqnarray}%
where it has been introduced the short-hand notation $\widehat{g}$ for the
scalar Kaluza-Klein fields $\widehat{g}_{ab}\left( x^{\mu }\right) $ and $R_{%
\mathbf{D}}$ is the Ricci scalar of the extra-dimensional manifold%
\begin{eqnarray*}
R_{\mathbf{D}} &=&-\widehat{g}^{ij}\left( C_{ai}^{k}C_{kj}^{a}+\frac{1}{2}%
C_{li}^{k}C_{kj}^{l}\right) -\widehat{g}^{mn}C_{im}^{i}C_{jn}^{j}+ \\
&&-\frac{\widehat{g}_{ij}\widehat{g}^{kp}\widehat{g}^{mn}}{4}%
C_{km}^{i}C_{pn}^{j}.
\end{eqnarray*}

It is apparent the origin of the UV divergences (mentioned in the
introduction) of the Kaluza-Klein Lagrangian for gauge and scalar fields (in
which the four dimensional part of the gravitational field is considered as
a classical background). The two most dangerous sources of
non-renormalizable interactions are the determinant ${\sqrt{g_{4+\mathbf{D}}}%
}$ of the metric in the gravitational action\footnote{%
The presence of such term ($\sqrt{g_{4+\mathbf{D}}}$, when $\rho $ is small,
is proportional to a constant plus $\rho $) generates non-renormalizable
interactions in which $\rho $ multiplies the kinetic terms of the $%
\overrightarrow{\pi }$ scalar fields (defined in Eq. (\ref{pidef})) and of
the gauge fields.} and the term 
\begin{equation*}
\frac{g^{\mu \nu }g^{\alpha \beta }}{4}\widehat{g}_{ab}F_{\mu \alpha
}^{a}F_{\nu \beta }^{b}
\end{equation*}%
in the Ricci scalar. Such a term also generates non-renormalizable
interactions between the $\rho $ field and the gauge fields as well as a
non-renormalizable interactions between the $\overrightarrow{\pi }$ fields
(defined below in Eq. (\ref{pidef})) and the gauge fields\footnote{%
The reason is that when one expands $\widehat{g}_{ab}$ around the chosen
ground state the expansion contains a term proportional to $\rho \delta
_{ab} $ as well as a term proportional to $\overrightarrow{\pi }\cdot 
\overrightarrow{t}$ (where $\overrightarrow{\pi }$\ are defined in Eq. (\ref%
{pidef})\ and the $\overrightarrow{t}$ are the generators of the algebra of $%
SO(\mathbf{D})$ in the tensor product of the adjoint representation with
itself).}. For the reasons mentioned at the beginning of the next section,
it is not possible to give a complete treatment of the renormalization of
the above Kaluza-Klein action. However, it is interesting to stress that the
one of the main guilties of the UV problems is the scalar degrees of freedom 
$\rho $. If one would find a sound mechanism to suppress the propagator of $%
\rho $ one would also soften many of the UV divergences of the theory. We
will come back on this important point in the following.

It is convenient the following decomposition of the scalar fields:%
\begin{equation}
\widehat{g}=\exp \left( 2\hat{\rho}\mathbf{1}\right) \exp \left( 
\overrightarrow{\pi }\cdot \overrightarrow{t}\right)  \label{pidef}
\end{equation}%
where $\mathbf{1}$ is the identity and $\overrightarrow{t}$ are the
generators of the algebra $SO(\mathbf{D})$ in the tensor product of the
adjoint representation with itself, the matrix $\widehat{g}$ has been
decomposed into a factor belonging to the group $SO(\mathbf{D})$ and its
determinant $\exp \left( 2\hat{\rho}\mathbf{1}\right) $. Thus, in the ground
state both $\hat{\rho}$ and the $\overrightarrow{\pi }$ vanish so that $%
\widehat{g}_{ab}=\delta _{ab}$. The fields $\overrightarrow{\pi }$
correspond to fluctuations which leave the determinant of $\widehat{g}$\
unchanged while the field $\hat{\rho}$ corresponds to fluctuations of the
determinant of $\widehat{g}$. The fields $\overrightarrow{\pi }$ belong to
the algebra of $SO(\mathbf{D})$ and have two indices in the adjoint
representation so that in the 't Hooft notation will be represented by four
lines while the field $\hat{\rho}$ is a singlet under $SO(\mathbf{D})$.

It is worth noting here that $\hat{\rho}$ is precisely the analogue of the
scalar degree of freedom of the Abelian Kaluza-Klein framework (in which the
reduction from five to four dimensions is considered). In particular, this
implies that the extra scalar constraint which prevents one from having both
a constant scalar field and the usual Yang-Mills equations for the gauge
fields in the non-Abelian Kaluza-Klein framework is related to $\hat{\rho}$.

In order to assure a proper behavior of determinant of $\widehat{g}$ in the
large $\mathbf{D}$ limit, the $\hat{\rho}$ field will be normalized as
follows:

\begin{equation}
\rho =\frac{2}{\mathbf{D}(\mathbf{D}-1)}\hat{\rho},  \label{rhodef}
\end{equation}%
(where $\rho $ is a field which is finite in the large $\mathbf{D}$ limit)
since, in this way, when $\mathbf{D}\rightarrow \infty $ the determinant of $%
\widehat{g}$ stays finite.

In terms of these fields, the kinetic terms of the scalars read%
\begin{equation*}
-tr\left(\left(\nabla^{\mu}\overrightarrow{\pi}\cdot\overrightarrow{t}%
\right)\left(\nabla_{\mu}\overrightarrow{\pi}\cdot\overrightarrow{t}%
\right)\right)+\left(1-\frac{{4}}{{\left(\mathbf{D}-1\right)^{2}\mathbf{D}%
^{2}}}\right)\nabla_{\mu}\rho\nabla^{\mu}\rho
\end{equation*}
where one recognizes, except by a constant factor, the usual kinetic terms
for scalar fields.

The large $\mathbf{D}$ scaling suggests that the scalar mode $\rho $ is
sub-dominant with respect to the Kaluza-Klein gauge fields and $%
\overrightarrow{\pi }$ scalars whose numbers grow with $\mathbf{D}$. This
suggests that the well known problem which arises in Kaluza-Klein scenarios
when one tries to deal with non-trivial Kaluza-Klein gauge fields but with
constant scalars arises at order $1/\mathbf{D}$. As we shall explain in the
next section, large $\mathbf{D}$ effects strongly suppress the propagator of
the $\rho $ field.

\section{Some examples of non-trivial Large $\mathbf{D}$ resummations}

We will now describe some non-trivial features of the large \textbf{$D$}
expansion of the Kaluza-Klein gauge and scalar fields. The 't Hooft
expansion in the Kaluza-Klein Lagrangian presents novel features due to the
appearance of scalar fields represented by four internal lines in the usual
"{}large \textbf{N}"{} notation. This lead to resummation which soften the
UV problem of the theory in a quite systematic way.

We will not try in the present paper to prove the full renormalizability of
the Kaluza-Klein Lagrangian for scalars and gauge fields. As it is well
known, for theory with gauge symmetry, the powerful methods of algebraic
renormalization based on the BRST symmetry have been developed (for a
detailed book on these methods see \cite{PiSo}). These tools allow one to
prove the quantum consistency of the BRST symmetry to all order in the gauge
coupling constant: the proof is recursive in the coupling constant itself.
However, these techniques cannot be applied in the large $\mathbf{D}$
expansion of the Kaluza-Klein Lagrangian: the reason is that in the usual
case of Yang-Mills theory the dependence of the classical action plus the
gauge fixing term on the coupling constant is very simple (a polynomial).
While, in the present case, the dependence of the Kaluza-Klein action on the
coupling constant $1/\mathbf{D}$ is quite complicated and very far from
being a simple polynomial: this prevents one from using the techniques of 
\cite{PiSo}\ in the present case\footnote{%
To the best of the authors knowledge, the renormalization procedure in the
large \textbf{N} expansion has been only developed for theories with global
symmetry (such as the Gross-Neveu model in three dimensions which is
renormalizable at large \textbf{N} despite being non-renormalizable in the
usual perturbative expansion). When dealing with the large \textbf{N}
expansion of QCD one does not worry about the renormalizability of the
theory at large \textbf{N} since the theory is already known to be
renormalizable by other means. Indeed, (unlike the cases with global
symmetries like the Gross-Neveu model) cases of gauge theories which are not
renormalizable in the usual perturbative expansion but can be renormalized
in the large \textbf{N} expansion are not known.}.

For these reasons, we will content ourself by showing that the large $%
\mathbf{D}$ expansion\ leads to a surprising softening of the UV divergences
by considering two simple examples: since we are considering the UV limit,
when computing the propagators and vertices the background geometry will be
assumed to be flat.

\subsection{Examples of large $\mathbf{D}$ corrections to the Scalar
propagators}

It will be now discussed the simplest correction to the $\rho $ propagator
due to Kaluza-Klein gauge fields loops. It will be considered the expansion
of $\widehat{g}_{ab}$ (defined in terms of the fundamental fields $\rho $
and $\overrightarrow{\pi }$\ in Eq. (\ref{pidef})) around the natural
groundstate $\left. \widehat{g}_{ab}\right\vert _{GS}$:%
\begin{equation*}
\left. \widehat{g}_{ab}\right\vert _{GS}=\delta _{ab}
\end{equation*}%
in such a way that the Feynman rules for the fundamental fields $\rho $, $%
\overrightarrow{\pi }$ and the gauge fields can be read directly from the
Lagrangian Eq. (\ref{larged1}).


In what follows, we will only need the bare propagators of $\rho $, $A$ and $%
\overrightarrow{\pi }$ which read:%
\begin{eqnarray*}
\Pi _{\rho }(k) &=&\left( 1-\frac{{4}}{{\left( \mathbf{D}-1\right) ^{2}%
\mathbf{D}^{2}}}\right) ^{-1}\frac{i}{k^{2}} \\
\Pi _{\pi }^{abcd}(k) &=&\frac{i\delta ^{ab}\delta ^{cd}}{k^{2}} \\
\Pi _{A\mu \nu }^{ab}(k) &=&\frac{-i\delta ^{ab}g_{\mu \nu }}{k^{2}}
\end{eqnarray*}%
where the $A$ propagator is in the
Feynman gauge and $k$ is the 4-momentum of the particle. 

Here, we will only focus on the analysis of the vertices which do not appear
in the usual Yang-Mills theory: the ones coming from the $\widehat{g}%
_{ab}F_{\mu \alpha }^{a}F_{\nu \beta }^{b}$ term when expanding $\widehat{g}%
_{ab}$ around the ground state. Such vertices describe non-renormalizable
interaction in normal perturbation theory and its presence could be viewed
as problematic in the usual scheme. Nevertheless, the large $\mathbf{D}$
expansion leads to a surprising improvement, as we will see in the case of
the "$\rho AA$" vertex.

We are going to consider the contribution of such vertex to the $\rho $
propagator. The Feynman rule for the non-renormalizable vertex "$\rho AA$"
(which originates from the term $\widehat{g}_{ab}F_{\mu \alpha }^{a}F_{\nu
\beta }^{b}$ of the Lagrangian) is:

%
\begin{equation}
-i\delta ^{ab}\kappa p_{1}\cdot p_{2}\left( g_{\mu \nu }-\frac{p_{1\mu
}p_{2\nu }}{p_{1}\cdot p_{2}}\right)  \label{eq:RAA}
\end{equation}%
where $\kappa $ is the Newton's constant\footnote{%
As it has been already stressed, we are interested in the UV limit of the
theory. Therefore, when writing the Feynman rules, we will consider a flat
background geometry.}, $p_{1}$ and $p_{2}$ are the 4-momenta of the gauge
fields. 
.

\begin{figure}[tbp]
\begin{centering}
\includegraphics[scale=0.4]{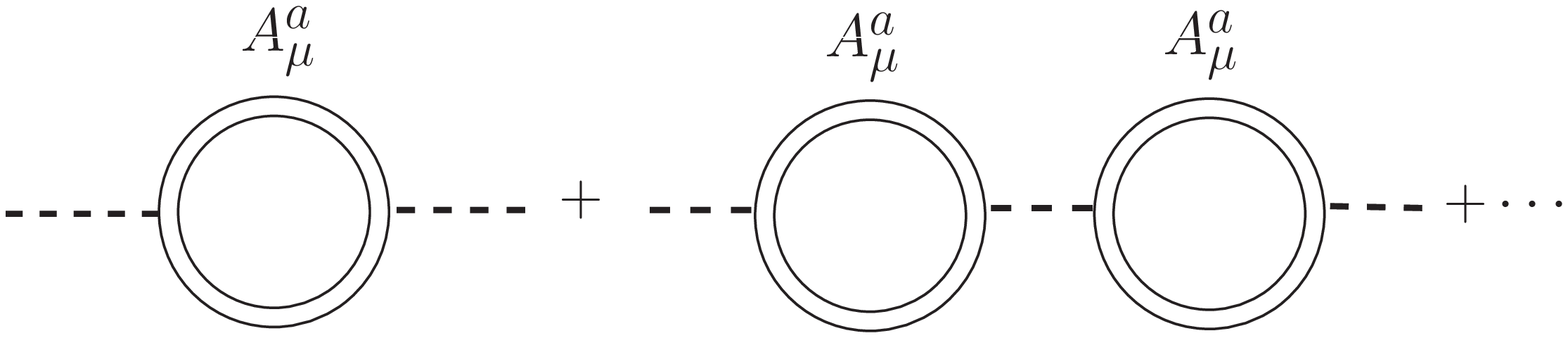}
\par\end{centering}
\caption{Corrections to the $\protect\rho$ propagator due to gluon loops}
\label{Flo:LoopCorrection}
\end{figure}

With this propagator, we can construct loop corrections to the $\rho$
propagator as shown in figure \ref{Flo:LoopCorrection} .

\begin{figure}[tbp]
\begin{centering}
\includegraphics[scale=0.4]{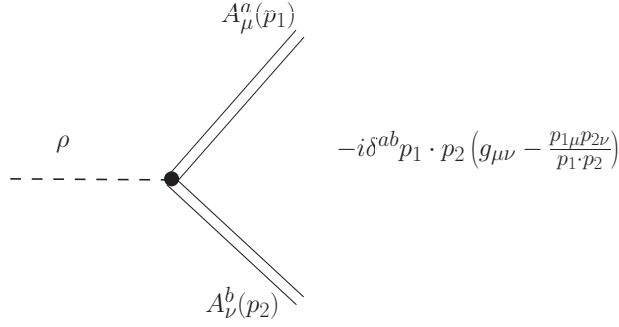}
\par\end{centering}
\caption{Feynman rule for the $\protect\rho A_{\protect\mu}^{a}A_{\protect\nu%
}^{b}$interaction }
\end{figure}

Each loop contributes with a term given by:

\begin{equation*}
\Delta _{\rho }\left( p\right) =\mathbf{D}(\mathbf{D}-1)\int d^{4}k\left\{ 
\frac{2\left[ k\cdot (p-k)\right] ^{2}+k^{2}\cdot \left( p-k\right) ^{2}}{%
k^{2}\cdot \left( p-k\right) ^{2}}\right\} 
\end{equation*}%
where $p$ is the 4-momentum of the $\rho $, $k$ is the internal 4-momentum
running in the loop and the dot represents the usual Lorentz product. Notice
that the integral is highly divergent but it can be regularized by the usual
methods. The presence of $\mathbf{D}(\mathbf{D}-1)$\ in the expression for $%
\Delta _{\rho }\left( p\right) $ can be easily understood by using the 't
Hooft notation.

When we sum up all the terms we obtain the geometric series and the result
reads:

\begin{equation*}
\left[\left( 1-\frac{{4}}{{\left( \mathbf{D}-1\right)
        ^{2}\mathbf{D}^{2}}}\right)p^{2}-\Delta _{\rho }\left(
    p\right)\right]^{-1} 
\end{equation*}

Because $\Delta _{\rho }$ is proportional to $\mathbf{D}(\mathbf{D}-1)$, we
find that the propagator of $\rho $ is suppressed in the large $\mathbf{D}$
limit strongly suggesting the decoupling of this degree of freedom in the
UV. This is an interesting effect since, in this way, all the
"non-renormalizable" loops in which the $\rho $ field appears are suppressed
as well by such large $\mathbf{D}$\ resummation. Therefore, this "large $%
\mathbf{D}$ resummation" leads to a clear improvement of the UV behavior of
the theory.

In a similar way, one can compute the correction to the propagator of the $%
\pi ^{ab}$ fields due to gauge fields loops. The Feynman rule for the
non-renormalizable vertex "$\pi AA$", which also originates from the term $%
\widehat{g}_{ab}F_{\mu \alpha }^{a}F_{\nu \beta }^{b}$\ of the Lagrangian as
shown in figure \ref{Flo:PAA}, is:


\begin{equation}
-i\delta ^{ac}\delta ^{bd}\kappa p\cdot p_{2}\left(g_{\mu \nu }-\frac{%
p_{1\mu }p_{2\nu }}{p_{1}\cdot p_{2}}\right)  \label{eq:PAA}
\end{equation}%
where,again, $p_{1}$ and $p_{2}$ are the 4-momenta of the gauge fields.

\begin{figure}[tbp]
\begin{centering}
\includegraphics[scale=0.4]{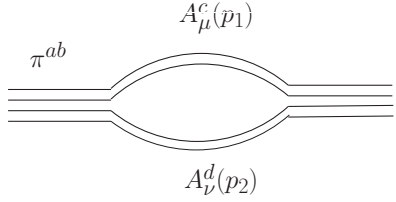}\label{Flo:Pi_loop}
\par\end{centering}
\caption{First correction to the $\protect\pi^{ab}$propagator}
\end{figure}

\begin{center}
\begin{figure}[tbp]
\begin{centering}
\includegraphics[scale=0.4]{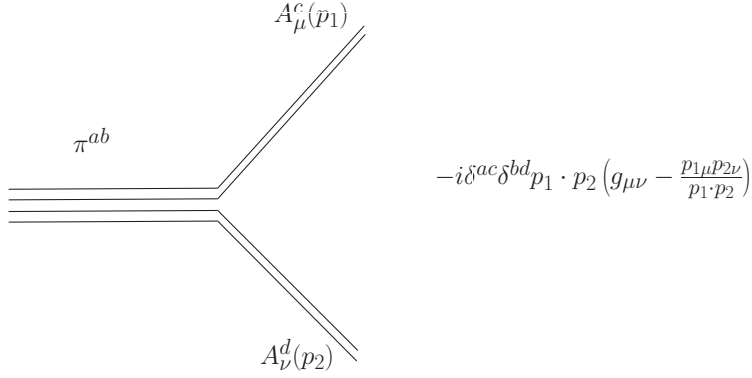}\label{Flo:PAA}
\par\end{centering}
\caption{Feynman rule for the$\protect\pi ^{ab}A_{\protect\mu }^{c}A_{%
\protect\nu }^{d}$ interaction}
\end{figure}

The main difference with the previous case is the {}\textquotedblleft color
index factor". Being $\pi ^{ab}$ is a {}\textquotedblleft four
line\textquotedblright\ field in t' Hooft diagrammatic notation (while the
gauge field is a usual {}\textquotedblleft two line\textquotedblright\
field), at the leading order there are not closed color lines in the loops
as we can in figure \ref{Flo:Pi_loop}.
\end{center}

Therefore, at the leading order each loop contribution is independent of $%
\mathbf{D}$:

\begin{equation*}
\Delta _{\overrightarrow{\pi }}\left( p\right) =2\int d^{4}k\left\{ \frac{2%
\left[ k\cdot (p-k)\right] ^{2}+k^{2}\cdot \left( p-k\right) ^{2}}{%
k^{2}\cdot \left( p-k\right) ^{2}}\right\} .
\end{equation*}%
Consequently, the $\overrightarrow{\pi }$ propagator is not directly
suppressed in the large $\mathbf{D}$ limit as in the case of $\rho $.

Eventually, since the ghosts fields must be in the same representation of the gauge group as the gauge fields, they will be represented in the 't Hooft notation by two internal lines. Consequently, it can be easily seen that they will "suffer" the same color corrections as the corresponding gauge fields: therefore the large $\mathbf{D}$ corrections to the ghosts propagators will be the same as the large $\mathbf{D}$ correction to the gauge fields. This should be enough to guarantee a consistent semi-classical expansion.

Indeed, in the present paper only two examples of "large $\mathbf{D}$"
corrections have been considered. However, the qualitative effects that one
should expect by including other possible vertices are consistent with the
effects discussed here: the reason is that in the large $\mathbf{D}$ limit
one can understand which are the dominant contributions by looking at the 't
Hooft double line notation\footnote{%
As a matter of fact, if one looks at the double line structure of the other
vertices, one can easily see, for instance, that generically large $\mathbf{D%
}$ effects tend to suppress the $\rho $ propagator because of the double
line structure shown in the above pictures.}. 

\section{Conclusions}

In the present paper some very interesting features of the large $\mathbf{D}$
expansion of a Kaluza-Klein compactification in $4+\mathbf{D}$ dimensions
have been analyzed. Firstly, it has been found that already at classical
level this model exhibits a nontrivial large $\mathbf{D}$ scaling: in
particular, it has been shown that the four dimensional effective
cosmological $\Lambda _{4}$ constant is of order $1/\mathbf{D}$ (so that, at
leading order in the large $\mathbf{D}$ expansion, $\Lambda _{4}$ vanishes)
whereas the size of the extra dimensions remains finite. At quantum level
some features of the 't Hooft large $\mathbf{D}$ expansion of the effective
Lagrangian for the scalar and gauge fields have been studied. It has been
shown that the scalar degree of freedom associated to the determinant of the
extra-dimensional metric (responsible of many UV divergences) is suppressed
in the large $\mathbf{D}$ limit: this effect strongly indicates that the UV
Kaluza-Klein divergences are softened in the large $\mathbf{D}$\ expansion.

As a final remark it is worth pointing out that
one could expect that this mechanism can work better than the usual
perturbative expansion even for a not extremely large value of
$D$. For instance in QCD already $N=3$ is enough to trust large $N$
expansion. On the other hand, it is well known that in order to
encompass the Standard Model within the Kaluza-Klein framework we need
at least seven extra dimensions.

\section*{Acknowledgements}

{\small {This work was supported by Fondecyt grant 11080056. The work of F.
C. has been partially supported by the Agenzia Spaziale Italiana. The work
of A. R. Z. is partially support by Fondecyt grant 1070880. The work
of A.G. is partially supported by UACh-DID grant S-2009-57. The Centro de
Estudios Cient\'{\i}ficos (CECS) is funded by the Chilean Government through
the Millennium Science Initiative and the Centers of Excellence Base
Financing Program of Conicyt. CECS is also supported by a group of private
companies which at present includes Antofagasta Minerals, Arauco, Empresas
CMPC, Indura, Naviera Ultragas and Telef\'{o}nica del Sur. 
}}

\end{document}